# MULTISTAGE CSR MICROBUNCHING GAIN DEVELOPMENT IN TRANSPORT OR RECIRCULATION ARCS

C. -Y. Tsai#, Department of Physics, Virginia Tech, Blacksburg, VA 24061, USA
D. Douglas, R. Li, and C. Tennant, Jefferson Lab, Newport News, VA 23606, USA

*Abstract*

Coherent synchrotron radiation (CSR) induced microbunching instability has been one of the most challenging issues in the design of modern accelerators. A linear Vlasov solver has been developed [1] and applied to investigate the physical processes of microbunching gain amplification for several example lattices [2]. In this paper, by further extending the concept of stage gain as proposed by Huang and Kim [3], we develop a method to characterize the microbunching development in terms of stage orders that allow the quantitative comparison of optics impacts on microbunching gain for different lattices. We find that the microbunching instability in our demonstrated arcs has a distinguishing feature of multistage amplification (e.g, up to 6th stage amplification for our example transport arcs, in contrast to two-stage amplification for a typical 4-dipole bunch compressor chicane). We also try to connect lattice optics pattern with the obtained stage gain functions by a physical interpretation. This Vlasov analysis is validated by ELEGANT [4] tracking results with excellent agreement.

## OVERVIEW OF CSR MICROBUNCHING INSTABILITY THEORY IN A SINGLE-PASS SYSTEM

Theoretical formulation of the CSR-induced microbunching instability in a single-pass system (e.g. a bunch compressor chicane) has been developed based on the linearized Vlasov equation [3, 5]. The formulation assumes initial modulation wavelength is small compared with the whole bunch duration (i.e. coasting-beam approximation) and treat the CSR effect as a small perturbation. By the method of characteristics, the equation that governs the evolution of the complex bunching factor can be written as [5]

$$b_k(s) = b_k^{(0)}(s) + \int_0^s K(s,s')b_k(s')ds' \quad (1)$$

where the bunching factor $b_k(s)$ is defined as the Fourier transform of the perturbed phase space distribution and the kernel function is particularly expressed as

$$K(s,s') = \frac{ik}{\gamma}\frac{I(s)}{I_A}C(s')R_{56}(s'\to s)Z\left(kC(s'),s'\right)\times[\text{Landau damping}] \quad (2)$$

for [Landau damping] term

$$[\text{Landau damping}] = \exp\left\{\frac{-k^2}{2}\left[\varepsilon_{x0}\left(\beta_{x0}R_{51}^2(s,s') + \frac{R_{52}^2(s,s')}{\beta_{x0}}\right) + \sigma_\delta^2 R_{56}^2(s,s')\right]\right\} \quad (3)$$

with

$$R_{56}(s'\to s) = R_{56}(s) - R_{56}(s') + R_{51}(s')R_{52}(s) - R_{51}(s)R_{52}(s') \quad (4)$$

and $R_{5i}(s,s') = C(s)R_{5i}(s) - C(s')R_{5i}(s')$.

Here the kernel function $K$(s,s') describes relevant collective effects, $g_k$(s) the resultant bunching factor as a function of the longitudinal position given a wavenumber $k$, and $g_k^{(0)}$(s) is the bunching factor in the absence of collective effect. $I(s)$ is the beam current at $s$ and $I_A$ is the Alfven current.

In this paper, we are interested in the bunching factor evolution subject to the CSR effect. For an ultrarelativistic electron beam traversing through a bending magnet, the CSR effect, described in terms of the impedance, can be expressed as [6, 7]

$$Z_{CSR}^{ss}\left(k(s);s\right) = \frac{-ik(s)^{1/3}A}{\left|\rho(s)\right|^{2/3}},\ A \approx -0.94 + 1.63i \quad (5)$$

where $k = 2\pi/\lambda$ is the modulation wave number, $\rho$ is the bending radius.

Here we presumed the CSR interaction be in the steady state and only in the longitudinal direction with negligible shielding effect. So far we have obtained the governing equation for the bunching factor and given the 1-D steady-state ultrarelativistic CSR impedance. In the following two sections, we would introduce two methods to solve Eq. (1), i.e. the direct solution and iterative solution, and define the microbunching gain functions associated with the two kinds of solutions, respectively, for our subsequent analysis.

## DIRECT SOLUTION

Here by "direct solution" we mean self-consistent solution of Eq. (1), as summarized below. First, we re-write Eq. (1) by expressing the bunching factors in vector forms and the kernel function in a matrix form, and we have after taking the inverse on both sides,

$$\mathbf{b}_k = \left(\mathbf{I}-\mathbf{K}\right)^{-1}\mathbf{b}_k^{(0)} \quad (6)$$

provided the inverse matrix of (**I**-**K**) exists.

To quantify the microbunching instability in a single-pass system, we define the microbunching gain as functions of the global longitudinal coordinate s as well as the initial modulation wavelength $\lambda$ (or, $k = 2\pi/\lambda$)

$$G(s,k=2\pi/\lambda) \equiv \left|\frac{b_k(s)}{b_k^{(0)}(0)}\right| \quad (7)$$

Hereafter, we simply call $G$(s) the gain function as a function of s given a specific modulation wavenumber, and denote $G_f(\lambda)$ gain spectrum as a function of $\lambda$ at a specific location (e.g. denoted with a subscript "$f$" at the exit of a beamline). Before ending this section, it deserves to mention the physical meaning of Eq. (1 or 6) and Eq. (7) with CSR effect [3]: a density perturbation at s' induced an energy modulation through CSR impedance

# jcytsai@vt.edu

and is subsequently converted into a further density modulation at s via momentum compaction function $R_{56}$.

## ITERATIVE SOLUTION

Another approach to solve Eq. (1) is resorted to iterative method, thus called iterative solution. Here we presume the zeroth order solution to be

$$\mathbf{b}_k^{(0)} = \mathbf{b}_k^{(0)} \tag{8}$$

and define the first order solution as

$$\mathbf{b}_k^{(1)} = (\mathbf{I}+\mathbf{K})\mathbf{b}_k^{(0)} \tag{9}$$

Then, the second order solution can be defined accordingly

$$\mathbf{b}_k^{(2)} = \left(\mathbf{I}+\mathbf{K}+\mathbf{K}^2\right)\mathbf{b}_k^{(0)} \tag{10}$$

In general, we have the n-th order solution to be expressed as

$$\mathbf{b}_k^{(n)} = \left(\sum_{m=0}^{n}\mathbf{K}^m\right)\mathbf{b}_k^{(0)} \tag{11}$$

It can be shown that Eq. (6) and Eq. (11) are equivalent when $n \longrightarrow \infty$, provided the sum converges. For a storage ring rather than a single-pass system, the convergence may not be held, which is however beyond the scope of this paper. We define the stage gain function with respect to Eq. (11) as follows

$$\tilde{G}^{(n)}(s,k=2\pi/\lambda) = \frac{\mathbf{b}_k^{(n)}(s)}{\mathbf{b}_k^{(0)}(0)}, \quad \text{and } G^{(n)}(s,k) = \left|\tilde{G}^{(n)}(s,k)\right| \tag{12}$$

We have mentioned the physical meaning of Eqs. (1) or (6) subject to CSR effect in the previous section. Here we give another interpretation by Eq. (11): the overall CSR gain at a specific position, say, at the exit of a lattice, can be contributed by many "staged gains." Let us take a 3-dipole bunch compressor chicane lattice as an example (see Fig. 1). The 0th-satge gain comes from pure optics effect [i.e. in the absence of collective effect, Eq. (8)]. The 1st-stage gain is contributed from initial density modulations (located at the beamline entrance, the first and/or second dipole entrance), converted to energy modulation via CSR interaction within the first and/or second dipole, then freely propagated by optics through $R_{56}$, to the last dipole via *one* interaction [second term on R.H.S. of Eq. (9)]. The 2nd-stage gain evolves from an initial density modulation (located at the beamline entrance or the first dipole), converted to energy modulation (via CSR within the first dipole) and then further density modulation (via $R_{56}$) till the second dipole, and such density modulation (which had experienced one-time CSR-$R_{56}$ conversion earlier) eventually turns into farther energy modulation via CSR within the second dipole and downstream $R_{56}$ till the last dipole, contributing to (part of) the resultant overall CSR gain [third term on R.H.S. of Eq. (10)]. To express in an alternative but more general way: the 1st-stage amplification refers to CSR interaction taking place inside only in *one* dipole (either 1st or 2nd dipole) where CSR impedance induces energy modulation as a result of density modulation. The microbunching structure in the beam evolves under optical propagation for the rest of the beamline. The 2nd-stage amplification refers to CSR interaction taking place inside *two* dipoles, with the beam phase space evolving under optical propagation for the rest of the beamline.

Figure 1 gives a conceptual diagram for the process to evolve. In this paper we consider multi-dipole system in a transport or recirculation arc lattice (e.g. Fig. 1d) in terms of multi-stage amplification scheme. In the following section of the stage gain analysis, we would quantify such multi-stage behavior of CSR microbunching gain in a general linear lattice.

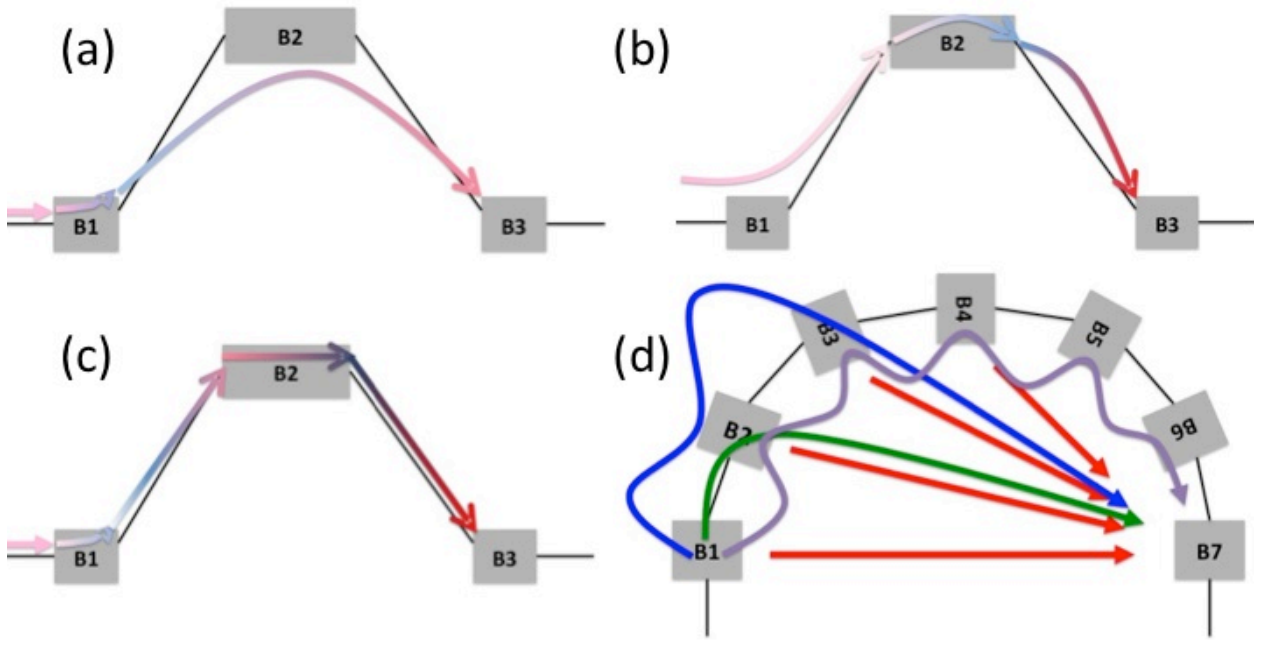

Figure 1: Conceptual illustration of multistage CSR microbunching gain evolution. For a typical 3- or 4-dipole bunch compressor chicane (a-c), (up to) 2-stage amplification can describe the microbunching gain evolution. Here for (a-c) the red color indicates the density modulation and the blue color represents energy modulation. Deeper colors indicate further amplified (or, more induced) modulations than for shallower colors.

## STAGE GAIN ANALYSIS

In this section, we intend to quantify the CSR gains by separating the contributions of beam parameters from the lattice properties and to extract individual stage gains from the overall CSR gain. To achieve this, we expand Eq. (12) in a series of polynomials of the beam current $I_b$ up to a certain order $M$,

$$\tilde{G}_f^{(M)} = \tilde{G}^{(M)}(s=s_f) = \tilde{G}_0 + \tilde{G}_1 I_b + ... + \tilde{G}_M I_b^M = \sum_{m=0}^{M}\tilde{G}_m I_b^m \tag{13}$$

By inspecting the kernel function, Eq. (2), the above expression can be further formulated to be

$$\tilde{G}_f^{(M)} = \sum_{m=0}^{M} A^m d_m^{(\lambda)}\left(\frac{I_b}{\gamma I_A}\right)^m \tag{14}$$

where $A$ is given in Eq. (5), $\gamma$ is the relativistic factor and $d_m^{(\lambda)}$ is the dimensionless coefficient (given a certain modulation wavelength) which now reflects the properties from lattice optics at $m^{\text{th}}$ stage ($m$ = 0, 1, 2,…), as well as Landau damping through finite beam emittances and energy spread [Eq. (3)]. For our interest in the following discussion, $\lambda$ is chosen to correspond to the maximal CSR gain, denoted as $\lambda_{opt}$. Here we point out that Eq. (38) of Ref. [4] can be a special case of Eq. (14) for $M$ = 2 in a typical bunch compressor chicane.

Obtaining the coefficients $d_m^{(\lambda)}$ of Eq. (14) can be straightforward. Here we remark the close connection between Eq. (2) and Eqs. (11) and (12) for determination of $d_m^{(\lambda)}$. For now, we can define the *individual* stage gain, which shall be convenient for our further discussion,

$$\mathcal{G}_f^{(m)} = \left| A^m d_m^{(\lambda)} \left( \frac{I_b}{\gamma I_A} \right)^m \right| \tag{15}$$

In the following section, we would take two comparative example arc lattices to demonstrate the stage gain analysis and its connection to both direct and/or iterative solutions.

## EXAMPLES

In this section we take two 1.3 GeV high-energy transport arcs as our comparative examples (hereafter dubbed Example 1 and Example 2 lattice). The detailed description of the two example lattices can be found in Ref. [7]. Table 1 summarizes some initial beam parameters used in our simulations. Here, Example 1 lattice is a 180° arc with large momentum compaction ($R_{56}$), as well as a second-order achromat and being globally isochronous with a large dispersion modulation across the entire arc. In contrast to the first example, Example 2 is again a 180° arc with however small momentum compaction. This arc is also a second-order achromat but designed to be a locally isochronous lattice within superperiods. Local isochronisity ensures that the bunch length is kept the same at phase homologous CSR emission sites. The lattice design strategy was originally aimed for CSR-induced beam emittance suppression, while our simulation results show that it appears to work for microbunching gain suppression as well. Figure 2 shows the Twiss functions and transport functions $R_{56}$(s) (or, the momentum compaction functions) across the arcs. Note that $R_{56}$(s) for Example 2 (Fig. 2d) is much smaller in amplitude than that for Example 1 (Fig. 2c) due to local isochronicity.

Table 1: Initial beam and Twiss parameters for the two example arc lattices

| Name | Example 1 (large $R_{56}$) | Example 2 (small $R_{56}$) | Unit |
|---|---|---|---|
| Beam energy | 1.3 | 1.3 | GeV |
| Bunch current | 65.5 | 65.5 | A |
| Normalized emittance | 0.3 | 0.3 | μm |
| Initial beta function | 35.81 | 65.0 | m |
| Initial alpha function | 0 | 0 | |
| Energy spread (uncorrelated) | $1.23\times10^{-5}$ | $1.23\times10^{-5}$ | |

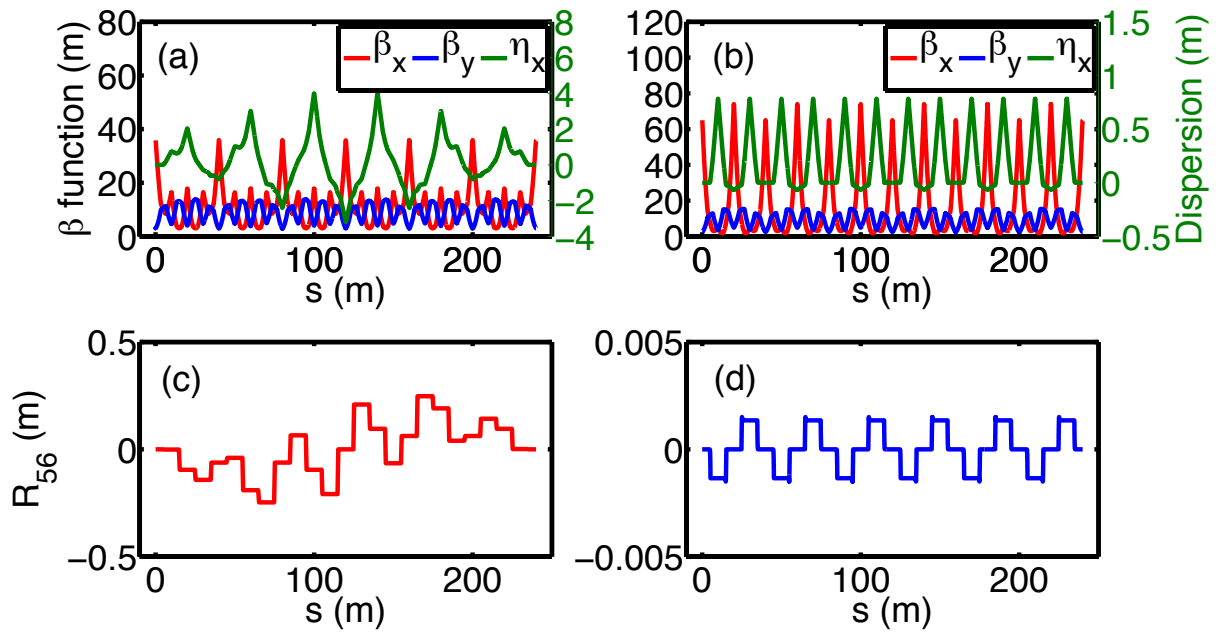

Figure 2: Lattice and transport functions for 1.3 GeV high-energy transport arc: (a)(c) with large momentum compaction function $R_{56}$ (Example 1); (b)(d) with small momentum compaction function $R_{56}$ (Example 2).

CSR microbunching gains for the two transport arcs are shown in Figs. 3 and 4. Figure 3 shows the gain spectra $G_f(\lambda)$ at the exits of the lattices as a function of modulation wavelength, from which one can obviously see a significant difference between them: Example 1 is vulnerable to CSR effect while the microbunching gain in Example 2 remains around unity. Figure 4 demonstrates the evolution of CSR microbunching gains as a function of s for several different wavelengths. One can see, in Fig. 3, that the shorter wavelengths enhance the Landau damping through Eq. (3), while longer wavelengths feature negligible CSR effect.

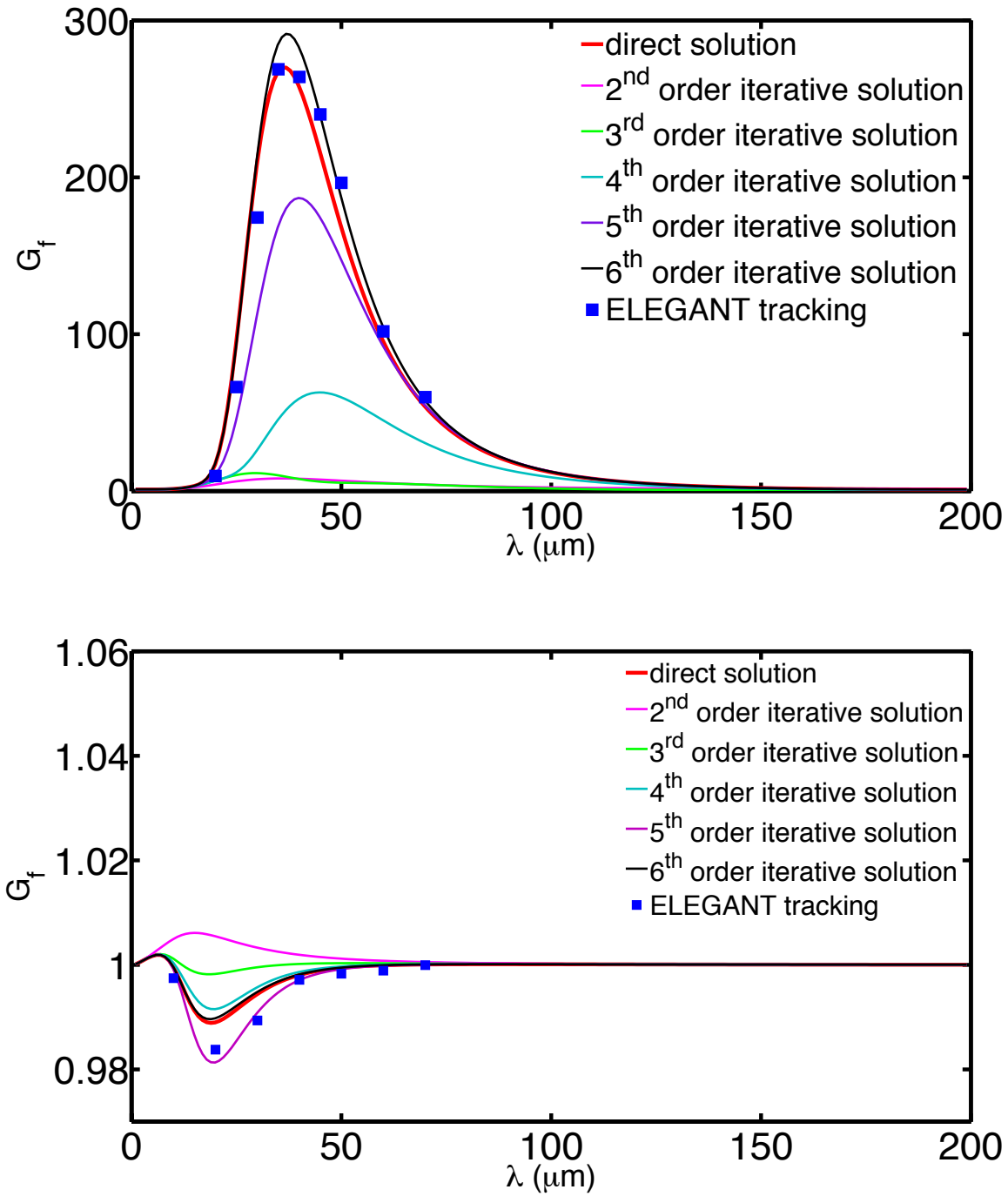

Figure 3: CSR gain spectra $G_f(\lambda)$ as a function of initial modulation wavelength for Example 1 (top) and 2 (bottom) lattice. The iteration solutions are obtained by Eq. (12).

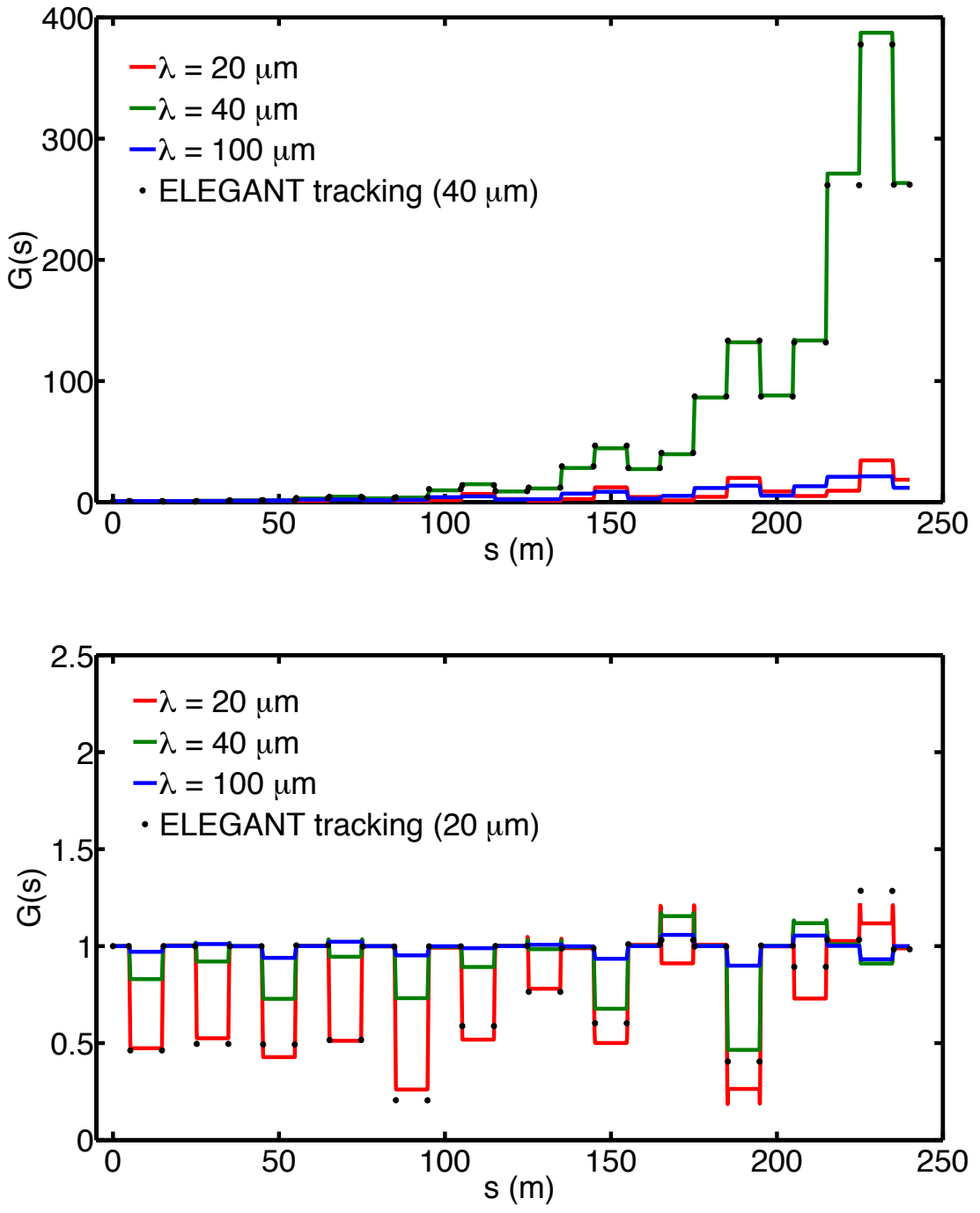

Figure 4: CSR gain functions $G(s)$ for Example 1 and 2 lattice.

From the simulation results (Figs. 3 and 4), we conclude that different lattice optics can give dramatically different CSR microbunching gains, although the geometric layout of the two lattices is identical. Also, we observe an interesting phenomenon: the two transport arcs are characteristic of (up to) 6th stage gain, which is distinguished from the (up to) 2nd-stage gain in a bunch compressor chicane [3]. Now, we would like to look into the gain amplification (or, gain evolution) in further depth by raising the following two questions: (i) how does CSR gain evolve along the beamline; based on the stage gain concept, can we quantify the CSR gain for each individual stages? (ii) Any advantage of employing the stage gain concept?

We still take Example 1 and 2 arcs as examples to extract the coefficients $d_m^{(\lambda)}$ [see Eq. (14)] so that we can quantify and compare optics impacts on the microbunching gains due to CSR interaction. Here we choose the (optimum) wavelengths 36.82 μm and 19 μm for Example 1 and 2, respectively. Figure 5 illustrates and compares the stage gain coefficients for the two arcs. Here we can see the coefficients for Example 1 are at least three orders of magnitude larger than those for Example 2, showing the essential difference in $d_m^{(\lambda)}$ between the two arcs. The dramatic difference of CSR microbunching gain for the two Example arcs can be attributed to the $d_m^{(\lambda)}$ difference. Figure 6 shows the bar charts representing the individual staged gains at lattice exits $\mathcal{G}_f^{(m)}$ [see Eq. (15)] as functions of beam current and stage index for both transport arcs. Here we have two observations in Fig. 6: first, given a specific stage order (say, $q$), as the beam current increases, $\mathcal{G}_f^{(q)}$ also increases; second, for the same beam current, as the stage order increases, it does not necessarily imply $\mathcal{G}_f^{(q)}$ increase accordingly. This is because the stage gain coefficient's behavior depends on the properties of a lattice itself.

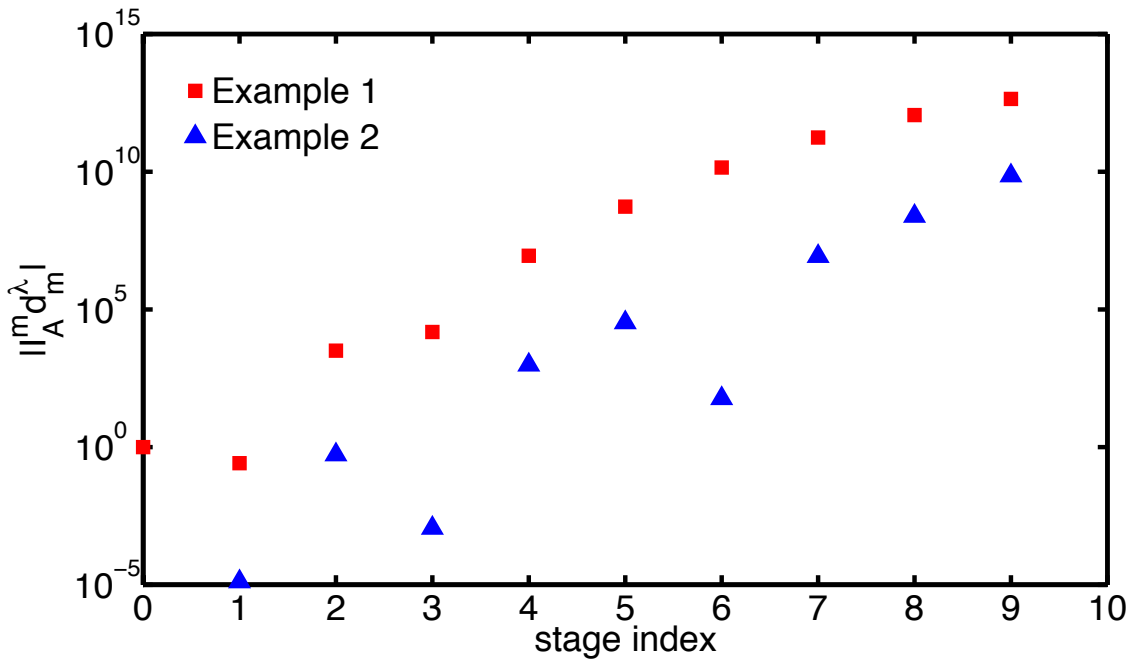

Figure 5: Comparison of $I_A^m d_m^{(\lambda)}$ for the two 1.3 GeV high-energy transport arcs; Example 1: red square and Example 2: blue triangle. Note the log scale in the vertical axis.

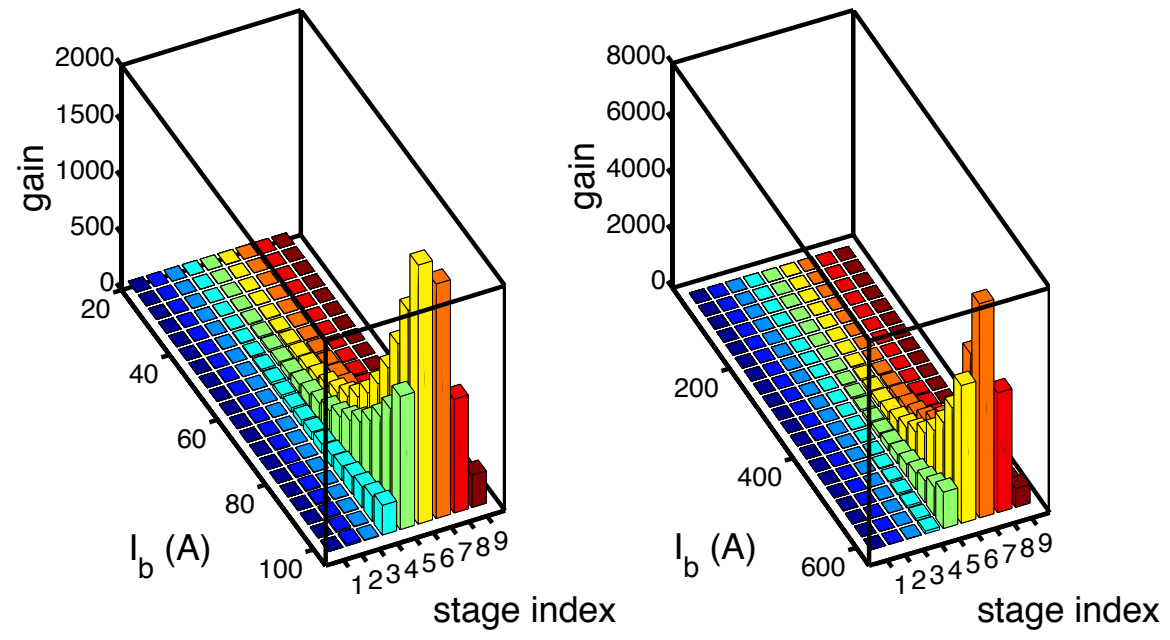

Figure 6: Bar chart representation of the individual staged gains [Eq. (18)] at the exits of the Example 1 and 2 lattices for several different beam currents. (Left) Example 1 ($\lambda$ = 36.82 μm); (right) Example 2 ($\lambda$ = 19 μm).

Regarding the advantage of the stage gain concept, since $d_m^{(\lambda)}$ is independent of beam current and beam energy, it can be used to quickly estimate the beam current dependence of the maximal CSR gain, provided an optimum wavelength is given. Figure 7 compares the current dependence of final overall gain from Eq. (14) for the two lattices at a selected wavelength that is in the vicinity of optimal wavelengths for maximal gains. It can be seen, in Example 2 case (Fig. 7b), the nominal beam current (65.5 A) is well described by including up to 6th order stage coefficient (red curve), while at further high currents (e.g. $I_b$ > 160 A), it needs to include higher stage orders into account (e.g. $M$ = 9, green curve). This observation is consistent with the 6th order iterative solutions presented in Fig. 3.

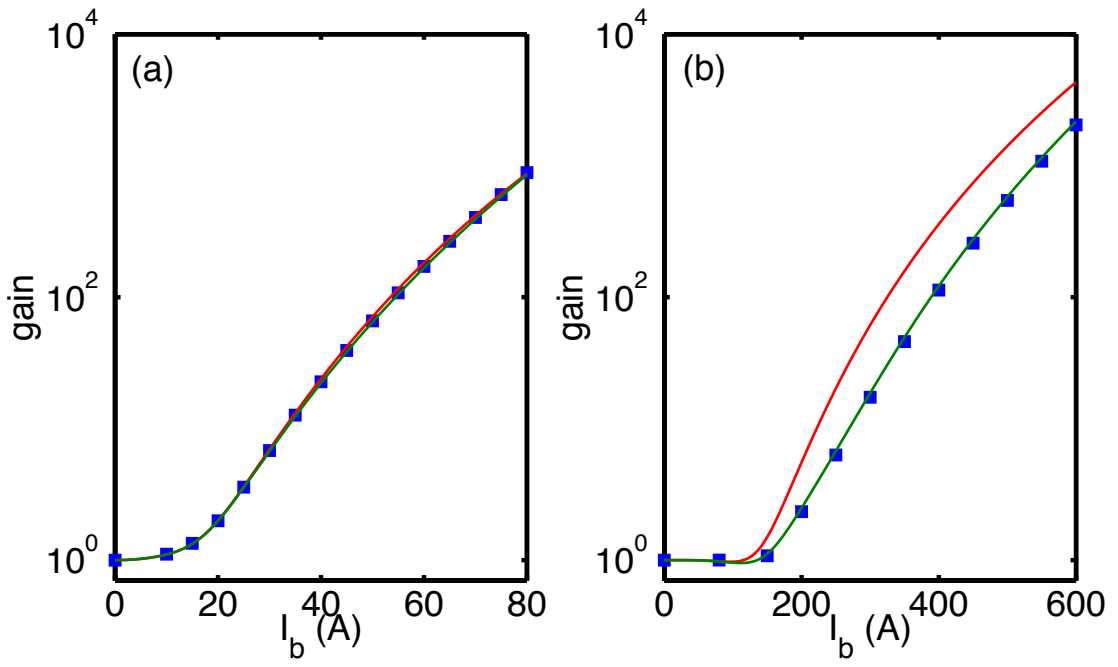

Figure 7: Current dependence of maximal CSR gain for the two high-energy transport arc lattices: (a) Example 1; (b) Example 2. Solid red curve from Eq. (14) with $M = 6$, solid green curve from Eq. (14) with $M$ = 9 and blue square dots from Eq. (7).

So far we have quantified the individual stage gains by extracting the coefficients $d_m^{(\lambda)}$ from the kernel function. The advantage of the extracted $d_m^{(\lambda)}$ has been used to make quick estimation of maximal CSR gains for a range of beam currents in a beamline. To answer our first question with our developed stage gain concept, it would be better to present $R_{56}(s' \to s)$ [defined in Eq. (4)] together in the analysis. Figure 8 shows the "quilt" pattern for the two example arcs. The upper left area in the figures vanishes due to causality. It is obvious that in Example 1 (left figure) those block areas with large amplitude, particularly the bottom right deep red blocks, can potentially accumulate the CSR gain. To be specific, for Example 1, energy modulation at $s'$ = 15 m can cause density modulation at $s$ = 60 m, where CSR can induce further energy modulation at the same location. Then such modulation propagates by $R_{56}(s' \to s)$ from $s'$ = 60 m to $s$ = 100 m, and so on. It is this situation that causes multi-stage CSR amplification. Here we note that more complete analysis needs to take Landau damping effect into account. In contrast, the situation for Example 2 (right figure) is more alleviated because of much smaller amplitudes in $R_{56}(s' \to s)$ . The microbunching amplification up to 6 stage in Example 1 and up to 9 stage in Example 2 are also manifested in Fig. 8.

Up to now, we have the above physical but qualitative interpretation of the multi-stage gain development along a beamline. We would like to more quantitatively connect the physical picture with our developed stage gain concepts. For simplicity, we exclude Landau damping effect and only consider the CSR microbunching amplification.

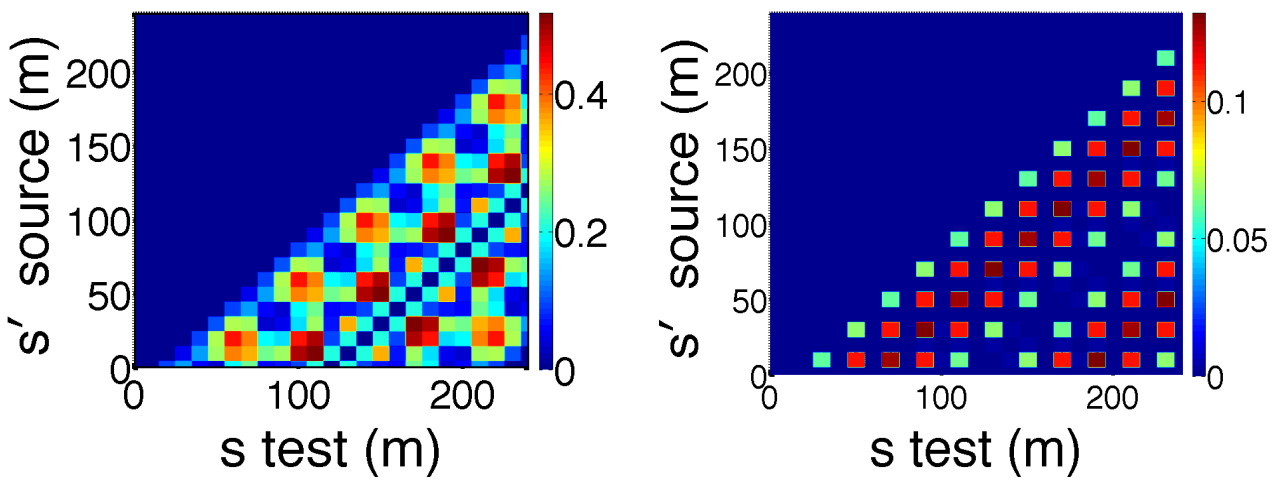

Figure 8: $R_{56}(s' \to s)$ quilt patterns for the two Example lattices: Example 1 (left) and Example 2 (right).

Figure 9 plots the staged gain functions $G^{(n)}(s)$ [defined in Eq. (12 or 14)] for Example 1 lattice without Landau damping effect [i.e. $\varepsilon_{nx} = \sigma_\delta = 0$ ], where we find the stage gain function is characteristic of periodic-like oscillation for lower-staged amplification (i.e. closely followed block patterns in left figure of Fig. 8) while features a stepwise increasing function for higher-staged amplification. It is this situation that reflects multi-stage CSR amplification. Similarly, for Example 2 lattice, there also exist many (even more) modular blocks (right figure of Fig. 8); however, in contrast to Example 1, the microbunching growth is less of a concern for Example 2 at a comparable bunch current (65.5 A) because of the smaller amplitudes of $R_{56}(s' \to s)$. The fact of even more modular blocks for Example 2 lattice would reflect its higher multi-stage gain behavior at higher currents, as can be seen in Fig. 7 (b). Note here that a higher stage does not correlated with higher amplitude of gain. Also it's important to remark that the staged-gain description in Eq. (11) has limited applications. For example, it is convergent only for a single-pass beamline when CSR interaction takes place in finite number of dipoles. Due to the same reason, multistage gain amplification concept as Eq. (11) may not be valid for longitudinal space charge (LSC) effect.

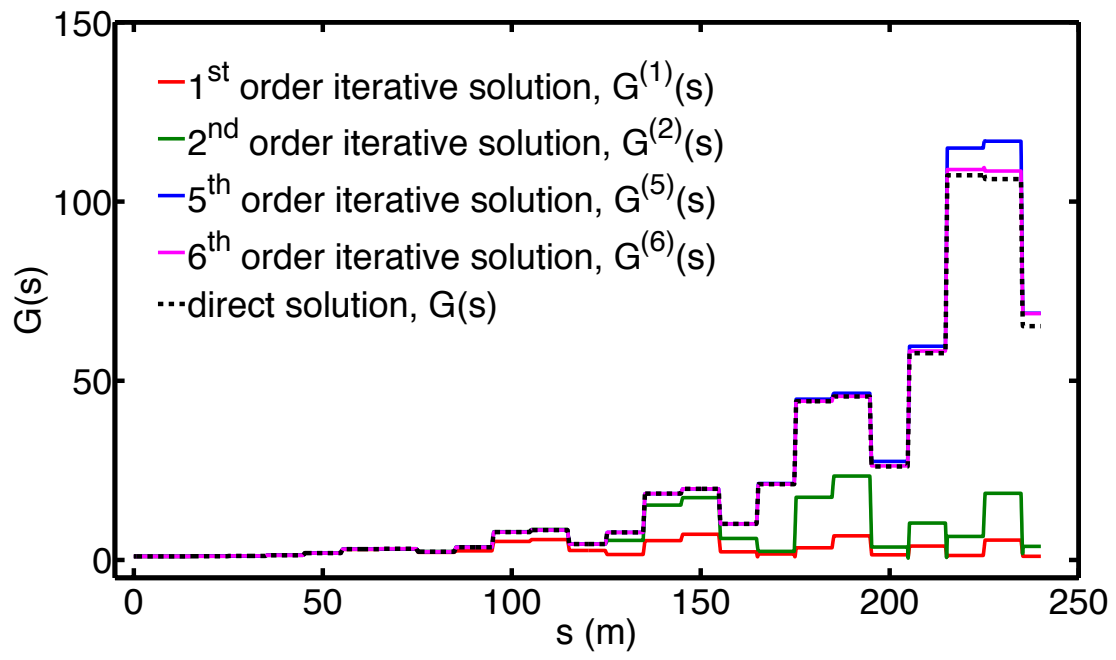

Figure 9: Gain functions $G^{(n)}$(s) (solid curves) and $G$(s) (dashed curves) for Example 1 lattice with $\lambda$ = 80 µm in the absence of Landau damping [Eq. (4)].

## CONCLUSION

In this paper, we have first outlined the theoretical formulation based on (linearized) Vlasov equation by treating the CSR effect as a perturbation and making the coasting beam approximation. The solution to the governing equation [Eq. (1)] can be obtain self-consistently (i.e. direct solution) or found through numerical iteration (i.e. iterative solution). With introduction of stage gain concept, the individual iterative solutions can be connected through the lattice optics pattern [i.e. $R_{56}(s' \to s)$ ] in a physical and quantitative way. Moreover, the stage gain coefficient [defined in Eq. (14)], due to its independence of beam current and beam energy, can be applied to make quick estimation for the maximal CSR gain, provided a lattice is given (Fig. 7).

## ACKNOWLEDGMENT

This work is supported by Jefferson Science Associates, LLC under U.S. DOE Contract No. DE-AC05-06OR23177.